# Contact Ion Pairs of Phosphate Groups in Water – Two-Dimensional Infrared Spectroscopy of Dimethyl-Phosphate and Ab-Initio Simulations

Jakob Schauss, Achintya Kundu, Benjamin P. Fingerhut, and Thomas Elsaesser[*]

*Max-Born-Institut für Nichtlineare Optik und Kurzzeitspektroskopie,*

*Berlin 12489, Germany*

*Corresponding author: elsasser@mbi-berlin.de

Abstract

The interaction of phosphate groups with ions in an aqueous environment has a strong impact on the structure and folding processes of DNA and RNA. The dynamic variety of ionic arrangements, including both contact pairs and water separated ions, and the molecular coupling mechanisms are far from being understood. In a combined experimental and theoretical approach, we address the properties of contact ion pairs of the prototypical system dimethyl-phosphate with $Na^+$, $Ca^{2+}$, and $Mg^{2+}$ ions in water. Linear and femtosecond two-dimensional infrared (2D-IR) spectroscopy of the asymmetric $(PO_2)^-$ stretching vibration separates and characterizes the different species via their blue-shifted vibrational signatures and 2D-IR lineshapes. Phosphate-magnesium contact pairs stand out as the most compact geometry while the contact pairs with $Ca^{2+}$ and $Na^+$ display a wider structural variation. Microscopic density functional theory simulations rationalize the observed frequency shifts and reveal distinct differences between the contact geometries.

TOC Graphic

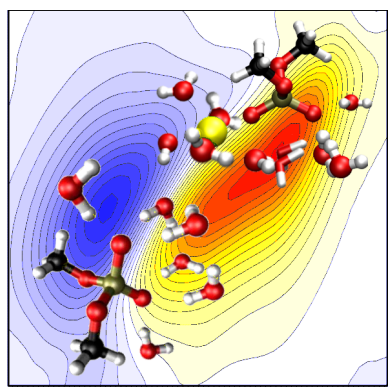

Contact pairs of oppositely charged ions represent a basic structural entity in ionic solutions, electrolytes, and biomolecular structures. The formation of contact ion pairs affects the electric conductivity of ionic solutions, thus providing access to the chemical equilibrium between contact pairs and ion species solvated separately (1-3). Structure research on hydrated DNA and RNA has identified contact ion pairs consisting of phosphate groups in the backbone and alkali and alkaline earth ions (4,5). Such on-site ion geometries are complemented by the so-called diffuse ion atmosphere which consists of mobile ions separated by at least one water layer from the surface of the biomolecule. The latter have remained largely elusive in x-ray diffraction patterns but a layer of enhanced ion concentration around helical DNA molecules has been observed in small angle x-ray scattering, suggesting a spatial 'condensation' of mobile ions due to the attractive Coulomb interaction with the helix (6,7,8). A clear distinction between on-site and water-solvated ions around DNA and RNA has, however, remained difficult and the role of the different ion geometries in stabilizing folded macromolecular structures is far from being understood. Moreover, insight in the dynamic exchange between the ion atmosphere and contact ion pair geometries which occurs in the fluctuating water environment, has remained very limited, calling for dynamic probes of local geometries and interactions at the molecular level (9).

Electrostatic interactions between ions in solution and counterion condensation have frequently been analyzed with the help of simulations based on the (linearized) Poisson-Boltzmann (PB) equation, applying a Coulomb interaction potential between ions embedded in a dielectric continuum (2,9,10). PB pictures have provided qualitative insight but have reproduced experimental results only in part. In particular, the neglect of the solvent molecular structure and local noncovalent interactions such as hydrogen bonds, the omission of electronic polarizabilities of the ions and the solvent, and the assumption of a static dielectric function are major shortcomings of PB pictures, making theory and simulation at the molecular level mandatory. Particular challenges are the treatment of polarizable ionic entities and water molecules, and, in molecular dynamics simulations, reliable force fields and sufficiently converged trajectories extending into the microsecond range (11,12,13).

The interaction of phosphate groups with alkali and alkaline earth ions in water has been studied via the stationary infrared and Raman spectra of phosphate vibrations in model systems for the DNA phosphodiester linkage such as diethyl- and dimethyl-phosphate (DMP) (14,15). Both the symmetric and asymmetric $(PO_2)^-$ stretching vibrations display frequency shifts under the influence of ions in their neighborhood. Extending such mostly qualitative studies, we have recently combined femtosecond two-dimensional infrared (2D-IR) spectroscopy and theoretical ab-initio simulations for a dynamic and quantitative characterization of contact ion pairs of DMP with magnesium ions ($Mg^{2+}$) (16). The 2D-IR spectra of DMP in water with an excess concentration of $Mg^{2+}$ display two distinct and uncoupled vibrational bands due to phosphate groups with and without an $Mg^{2+}$ ion in close contact. The band of the contact ion pairs is shifted to higher transition frequencies due to the repulsive exchange interaction (Pauli repulsion) between the $(PO_2)^-$ and the $Mg^{2+}$ entities, and the vibrational population decay is somewhat slowed down.

For a deeper understanding of contact ion pairs of phosphate groups and on-site ion geometries in DNA and RNA, it is important to systematically explore a wider range of mono- and divalent ions, including the most abundant sodium ($Na^+$) and calcium ($Ca^{2+}$) cations. The role of hydrating water molecules needs to be characterized by a combination of experiment and theory in order to get insight into particular atomic arrangements and their impact on the subtle interplay of the attractive Coulomb and the repulsive exchange interactions defining the spectroscopic signatures. In this Letter, we report an in-depth study of contact ion pairs of DMP with $Na^+$ and $Ca^{2+}$ ions, using the $Mg^{2+}$ results as a benchmark. The combination of linear infrared and 2D-IR spectroscopy with microscopic density functional theory simulations reveals distinct differences between the contact geometries and ensuing interactions, providing a systematic and quantitative picture of contact ion pairs containing phosphate groups.

In the experiments, solutions of the $Na^+DMP^-$ salt were prepared at a $c_0=0.2$ M concentration in ultrapure $H_2O$. An excess of $Na^+$, $Ca^{2+}$, or $Mg^{2+}$ ions was introduced by adding NaCl, $CaCl_2$, or $Mg(H_2O)_6Cl_2$ to this solution. The excess ion concentration was varied between 0.5 and 2.0 M. Details of the experimental procedures including the linear infrared and 2D-IR measurements are given in the Supporting Information (SI).

Linear infrared absorption spectra of the different samples are summarized in Fig. 1. The thick black line in Fig. 1(a) represents the absorption band of the asymmetric $(PO_2)^-$ stretching vibration $\nu_{AS}(PO_2)^-$ of DMP without ion excess. This band consists of two components with maxima at 1190 and 1215 cm$^{-1}$ which originate from the *gt* and *gg* conformers of DMP in solution (17). The main effect of $Na^+$ addition with a 2 M concentration consists in a decrease of absorption strength in the low-frequency part while spectral shifts are essentially absent. In contrast, addition of $Ca^{2+}$ and $Mg^{2+}$ ions with the same concentration leads, on top of the absorption decrease, to the occurrence of new blue-shifted absorption, clearly visible as a shoulder between 1230 and 1260 cm$^{-1}$.

In Fig. 1(b), differential absorption spectra are plotted for different excess concentrations of $Na^+$ (colored lines). These spectra were calculated by subtracting the stationary $\nu_{AS}(PO_2)^-$ band without excess ions (thick black line in panel (a)) from the respective spectrum of the samples with different ion excess concentrations $c_{ion}$. Both DMP molecules solvated by water and DMP molecules interacting with excess ions contribute to the differential spectra which exhibit a pronounced absorption decrease at low frequencies and a negligible absorption change above 1235 cm$^{-1}$. To extract the absorption of the DMP/ion complexes, we subtracted the scaled and inverted $\nu_{AS}(PO_2)^-$ band without excess ions (dash-dotted line in panel (b)) from the differential spectrum for $c_{ion}$ = 2M (blue line). This procedure is described in detail in the SI (equation (3) and related text). The dashed line in Fig. 1(b) represents the absorption band of DMP molecules interacting with $Na^+$ ions, displaying a moderate blue shift and a maximum around 1220 $cm^{-1}$. A similar behavior was observed with potassium ($K^+$) excess ions (not shown).

The impact of $Ca^{2+}$ and $Mg^{2+}$ excess ions on the $\nu_{AS}(PO_2)^-$ absorption spectra was studied under identical experimental conditions. The differential spectra shown in Figs. 1(c,d) demonstrate pronounced blue-shifted absorption components, on top of the respective absorption decrease at lower frequencies. The spectra of the DMP/ion complexes (dashed lines) consist of a strong blue-shifted absorption line and a weaker red-shifted contribution. The frequency up-shift of the main component relative to the maximum of the original DMP band at 1206 $cm^{-1}$ has a value of ~30 $cm^{-1}$ for $Ca^{2+}$ and ~40 $cm^{-1}$ for $Mg^{2+}$, both exceeding the ~20 $cm^{-1}$ blue-shift observed with $Na^+$ excess ions. Such differences point to a different impact of excess ions on the DMP vibrational potential, i.e., different interaction geometries and strengths for the different excess ion species. The blue-shifted component in the $Mg^{2+}$ spectra has been assigned to contact ion pairs of the $(PO_2)^-$ unit of DMP with $Mg^{2+}$ ions while the red-shifted component is due to solvent-separated pairs of DMP and $Mg^{2+}$ (16). First measurements with DNA and RNA duplexes and a controlled excess of $Na^+$, $Ca^{2+}$, and $Mg^{2+}$ reveal similar blue shifts of the $\nu_{AS}(PO_2)^-$ band upon contact ion pair formation. Linear infrared absorption spectra of such systems are shown in Fig. S2 of the SI.

The absolute values of differential absorbance in the spectra of Fig. 1 allow for estimating the concentration of DMP/ion complexes. The procedure applied to the spectra for a 2 M ion excess is described in the SI. In brief, minimum concentrations of the complexes are derived from the absorption spectra plotted as dashed lines in Fig. 1. Maximum concentrations are estimated from a comparison of the blue-shifted peak positions in the calculated linear absorption spectra of the complexes (Figs. 1 and S1) and the positions of the blue-shifted diagonal peaks in the 2D-IR spectra (diagonal cuts in Figs. 2 and S3). We find that the DMP/$Ca^{2+}$ concentration has a value between $0.12c_o$ and $0.3c_o$ while that of DMP/$Mg^{2+}$ contact ions is between $0.23c_0$ and $0.37c_0$ with $c_0=0.2$ M being the total concentration of DMP molecules. For DMP/Na⁺ complexes, we estimate a minimum concentration of $0.14c_0$. In other words, up to some 37% of the DMP molecules form contact ion pairs, while the fraction of excess ions in contact pairs is on the order of 1-4%. The uncertainty of these concentration values is on the order of 30%.

Nonlinear 2D-IR spectroscopy allows for a much better separation of the different spectral features than linear difference spectra and, moreover, gives insight into molecular couplings. In Fig. 2, we present a series of 2D-IR spectra measured for a population time T=300 fs with (a) DMP in neat water, (b) DMP and $Na^{2+}$ excess ions, (c) DMP and $Ca^{2+}$ excess ions, and (d) DMP and $Mg^{2+}$ excess ions. In all measurements, the DMP concentration was 0.2 M, the excess ions had a concentration of 2 M. The yellow-red and the blue contours in Fig. 2 show the absorptive 2D-IR signal on the v=0→1 and the v=1→2 transition of the DMP $\nu_{AS}(PO_2)^-$ vibration, plotted as a function of excitation and detection frequency. The first contribution is due to bleaching of the v=0 state and stimulated emission from the v=1 state, resulting in an absorption decrease, while the second contribution is due to the red-shifted excited state absorption from the v=1 state, corresponding to an absorption increase.

The 2D spectrum of DMP in neat $H_2O$ (Fig. 2a) displays a single peak on the v=0→1 transition with moderate inhomogeneous broadening, as is evident from the elliptic lineshape tilted with respect to the excitation frequency axis. A cut through this band along the frequency diagonal $\nu_1=\nu_3$ (Fig. 2d) exhibits a spectral width comparable to that of the linear absorption spectrum in Fig. 1(a). Cuts at constant $\nu_1$, i.e., parallel to the detection frequency axis, are given in SI (Fig. S4). Upon addition of Na⁺ excess ions, the v=0→1 feature of the 2D spectrum (Fig. 2b) broadens along the diagonal, as is evident from the diagonal cuts plotted in Fig. 2(f). Such behavior becomes very pronounced for $Ca^{2+}$ and $Mg^{2+}$ excess ions (Figs. 2g,h) where an additional blue-shifted component arises, corroborating the blue-shifted components in the linear spectra of Fig. 1. It is important to note that 2D cross peaks between such new contributions to the 2D spectra and the original DMP band are absent, i.e., the underlying vibrations are uncoupled and arise from distinct chemical species. With increasing population time T, the 2D signals display a first decay governed by the 300 – 600 fs lifetimes of the v=1 states, with slightly longer lifetimes of the blue-shifted components. This decay establishes a heated vibrational ground state, giving rise to small residual bleaching signals on the v=0→1 transitions. Femtosecond 2D-IR spectra for longer waiting times and pump-probe data are presented in the SI.

Microscopic insight into structural differences of the different contact ion pairs was obtained with the help of density functional theory simulations on DMP($H_2O$)$_N$$M^{x+}$ clusters ($M^{x+}$ = $Na^+$, $Mg^{2+}$: N = 11; $M^{x+}$ = $Ca^{2+}$: N = 13). The respective minimum geometries of the DMP($H_2O$)$_{11}$$Na^+$, DMP($H_2O$)$_{13}$$Ca^{2+}$ and DMP($H_2O$)$_{11}$$Mg^{2+}$ clusters are depicted in Figs. 3(a-c). DMP($H_2O$)$_N$$M^{x+}$ clusters were constructed as minimal models by saturating the first solvation shell around DMP and the ions with $H_2O$ molecules at the hydrogen bond acceptor positions (see SI for simulation details). Key geometric differences for the different ions were identified in the P…ion distance and the P…O1…ion angle α as relevant coordinates. The angle α takes a value of 180° for linear arrangements of the P=O group and the ion while α ≈ 90 ° if the ion is placed in the center of the bisector of the ($PO_2$)$^-$ group. The respective values of the angular coordinate α of the DMP($H_2O$)$_{11}$$M^{x+}$ clusters are indicated in Figure 3(a-c) and summarized in Table S1. We find that the prototypical solvation geometry realized in the DMP($H_2O$)$_{11}$$Na^+$ cluster is characterized by α ≈126°. $Na^+$ intercalates into the tetrahedral hydrogen bond geometry around the O1 atom of the phosphate group, replacing a single water molecule in the first hydration shell. For doubly charged $Ca^{2+}$, we find that the minimum of the DMP($H_2O$)$_{13}$$Ca^{2+}$ clusters is characterized by α≈152°, thus showing an approximate linear arrangement of the P=O group. In this prototypical solvation geometry the O1 atom of the ($PO_2$)$^-$ group takes the position of one of the water oxygen atoms in the first solvation geometry around the $Ca^{2+}$ ion. Constrained optimization in the minimal cluster model $DMP(H_2O)_{13}Ca^{2+}$ along the angular coordinate $\alpha = \sphericalangle(Ca^{2+}...O1...P)$ allows to interpolate between this prototypical solvation geometry and the tetrahedral hydrogen bond geometry around the O1 atom of the $(PO_2)^-$ group realized for $Na^+$, indicating a wider structural heterogeneity with minima of comparable energy (Table S1). Due to the rigid octahedral solvation shell of $Mg^{2+}(H_2O)_6$, a single water molecule of the $Mg^{2+}$ solvation shell is replaced by the O1 atom of the phosphate group in the prototypical solvation geometry of the $DMP(H_2O)_{11}Mg^{2+}$ cluster, thus preserving the octahedral geometry around $Mg^{2+}$. This geometry is characterized by an approximately linear arrangement of the P=O group and the counter ion ($\alpha \approx 173°$). Vibrational frequencies $\nu_{AS}(PO_2)^-$ of $DMP(H_2O)_{11}Na^+$, $DMP(H_2O)_{13}Ca^{2+}$ and $DMP(H_2O)_{11}Mg^{2+}$ are found to be 1208, 1226 and 1248 cm$^{-1}$, in close agreement to experiment.

Larger $DMP(H_2O)_{19}M^{x+}$ clusters ($M^{x+} = Na^+$, $Ca^{2+}$ and $Mg^{2+}$) were employed to get further insight into the observed blue shift of the $\nu_{AS}(PO_2)^-$ vibration (cf. Figs. 1 and 2). In the larger clusters, roughly two solvation shells can be formed between DMP and the ion even for P...ion distances up to 10 Å (cf. SI). Figure 4(a) presents the potential energy surfaces for a displacement along the asymmetric stretching vibration normal mode coordinate ($Q^{AS}_{PO2}$) for two selected P...$Mg^{2+}$ distances R (potential energy surfaces of the $DMP(H_2O)_NM^{x+}$ clusters with $M^{x+} = Na^+, Mg^{2+}$: N = 11 and $M^{x+} = Ca^{2+}$: N = 13 are given in SI, Fig. S8). For R = 10.088 Å the potential appears highly symmetric while for close contact between the phosphate group and $Mg^{2+}$ (R = 3.088 Å) deviations from the symmetric shape are clearly recognized. Notably, for positive displacements along $Q^{AS}_{PO2}$ (corresponding to P=O1 elongation) the potential is strongly destabilized due to the access to the repulsive part of the intermolecular DMP...$Mg^{2+}$ potential (cf. Fig. 3c).

Figure 4(b) presents the vibrational frequency of the asymmetric $(PO_2)^-$ stretching vibration (in harmonic approximation) as a function of P…ion distance R. The R-dependence of $Ca^{2+}$ closely resembles the reported values of $Mg^{2+}$ (16). For 10 Å > R > 6 Å the ions possess independent solvation shells, corresponding to solvent separated ion pairs that are separated by two $H_2O$ molecules. The dipolar field of the ions leads to an increased ordering of the water solvation shells, resulting in an increase of the electric field acting on the $(PO_2)^-$ group. The enhanced field induces a slight red-shift of the $(PO_2)^-$ stretching frequency (16). The red-shifted components in the linear infrared spectra of the DMP…$Ca^{2+}$ and DMP…$Mg^{2+}$ samples (dashed lines in Fig. 1c,d) point to the occurrence of such geometries with intermediate P…ion distances. For R < 4 Å contact ion pair formation leads to a blue shift of the asymmetric $(PO_2)^-$ stretching vibration that is less pronounced for $Ca^{2+}$ than for $Mg^{2+}$. For $Na^+$ the changes in vibrational frequency along R are smaller and within the experimental linewidth, only a slight blue-shift is observed upon contact ion pair formation with DMP.

Our combined experimental and theoretical results demonstrate that the $(PO_2)^-$ group of DMP forms contact ion pairs with $Na^+$, $Ca^{2+}$, and $Mg^{2+}$, however, with distinctly different geometric arrangements and interaction patterns. The structure of the contact ion pairs is determined by the interplay of attractive Coulomb and repulsive exchange interactions between the ionic constituents and by the arrangement of solvating water molecules. In parallel to the ions, the dipolar water molecules generate strong electric fields and interact via hydrogen bonds with the ion pairs (16,17,18,19). The DMP…$Mg^{2+}$ contact pairs display the most rigid structure in which one of the $(PO_2)^-$ oxygens replaces a water oxygen in the octahedral first solvation layer around $Mg^{2+}$. This results in a quasi-linear P=O…$Mg^{2+}$ arrangement with a particularly short P…$Mg^{2+}$ distance of 3.48 Å (O1…$Mg^{2+}$ distance ≈ 2.02 Å), and a blue-shift of the $(PO_2)^-$ stretching vibrations induced by the repulsive exchange interaction term in the vibrational potentials (16). NMR studies indicate that such geometries are preserved for a time range beyond 1 μs (20).

The water solvation shell of the DMP…$Ca^{2+}$ complexes (Fig. 3b) displays a larger structural variety where the quasi-linear P=O…$Ca^{2+}$ geometry found for $Mg^{2+}$ is largely preserved and the $(PO_2)^-$ oxygens replace a water oxygen in the first solvation layer around $Ca^{2+}$ (P…$Ca^{2+}$ distance ≈ 3.78 Å, O1… $Ca^{2+}$ distance ≈ 2.25 Å). The DMP…$Na^+$ complexes are the least rigid structure with an angled P=O…$Na^{2+}$ arrangement of contact ion pairs (Fig. 3a). In such an angled geometry the $Na^+$ ion intercalates into the tetrahedral hydrogen bond environment around the $(PO_2)^-$ oxygen atoms. For both DMP…$Ca^{2+}$ and DMP…$Na^{2+}$ contact ion pairs, the simulations predict a reduced blue-shift of the $\nu_{AS}(PO_2)^-$ absorption induced by Pauli repulsion, as observed in the experiment.

The water structure around DMP is different from the hydration geometries of phosphate groups in the backbones of DNA and RNA for which the helix geometries introduce additional structural boundary conditions (18,19,21,22). To explore contact ion pair formation in such systems, we have performed first experiments with double-stranded DNA and RNA oligomers as described in the SI. In presence of excess ions, the linear absorption spectra of the asymmetric phosphate stretch vibrations of DNA and RNA undergo changes very similar to those of DMP. There are pronounced blue-shifted absorption components in both the B-helix geometry of DNA and the A-helix geometry of RNA, pointing to the formation of contact ion pairs. Future work will address this behavior in a quantitative way and complement the linear absorption spectra by 2D-IR studies.

In conclusion, our comparative experimental and theoretical study of contact ion pairs of DMP with $Na^+$, $Ca^{2+}$, and $Mg^{2+}$ in water gives detailed insight in the relevant molecular interactions and resulting molecular structures, thus revealing for the first time characteristic structural differences between the contact pairs. The experimental results demonstrate that the asymmetric $(PO_2)^-$ stretching vibration represents a most sensitive probe of molecular structure and dynamics which can grasp subtle changes in coupling strengths and geometries. The contact ion pairs with $Mg^{2+}$ stand out with their particularly rigid, long-lived structure and the shortest distance between one of the $(PO_2)^-$ oxygens and the magnesium ions. This geometry strongly affects the electronic structure of the phosphate group, as manifested here in changes of the vibrational potential energy surface. Such observations support the particular relevance of magnesium ions in stabilizing folded biomolecular structures by modifying interaction patterns between charged molecular units.

*Acknowledgments.* This research has received funding from the European Research Council (ERC) under the European Union's Horizon 2020 research and innovation program (grant agreements No. 833365 and No. 802817). B. P. F. acknowledges support by the DFG within the Emmy-Noether Program (Grant No. FI 2024/1-1). We thank Janett Feickert for expert technical support.

*Supporting Information Available.* Experimental methods, experimental results, theoretical methods.

**Figures**

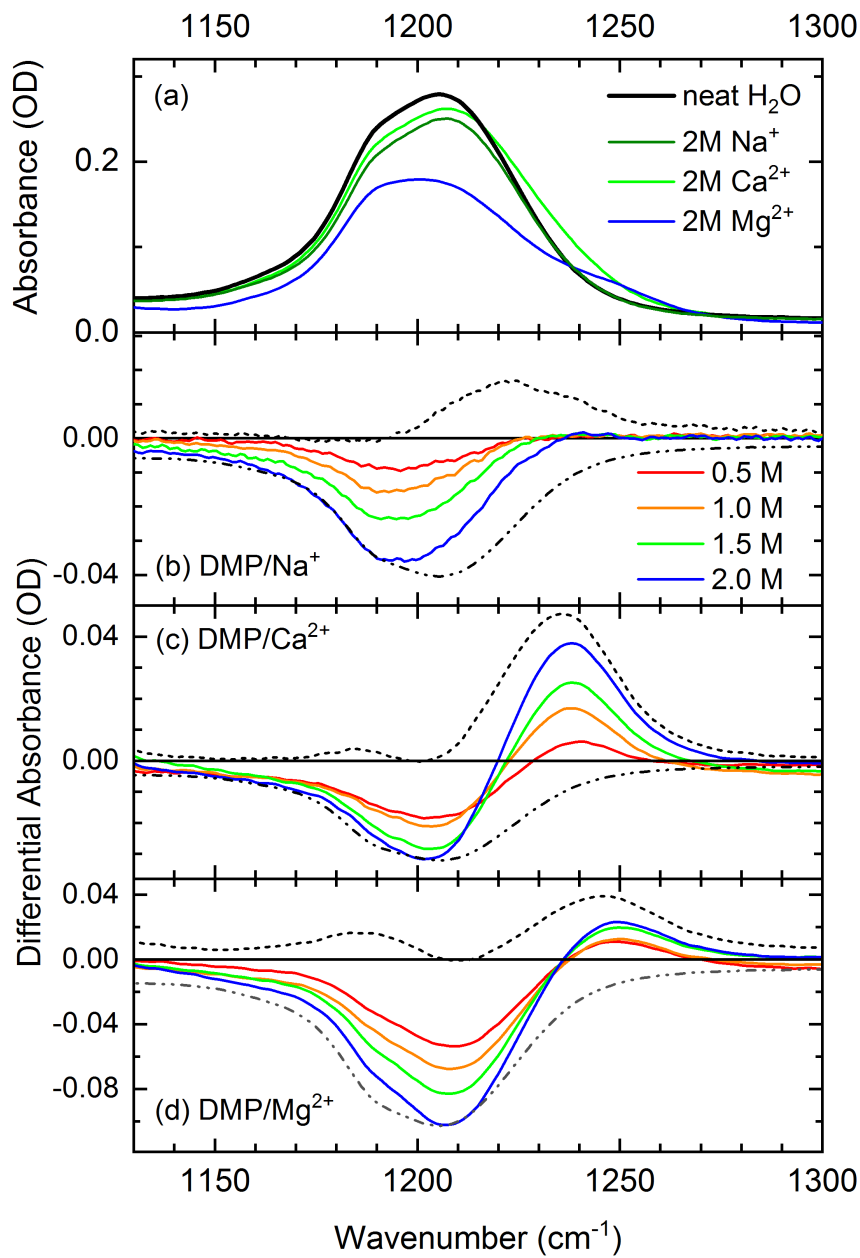

**Figure 1**. Linear infrared absorption band of the asymmetric phosphate stretch vibration of dimethylphosphate (DMP, total concentration $c_0$ = 0.2 M, sample thickness 25 μm). (a) DMP in neat water (thick black line) and DMP in aqueous solutions containing 2M NaCl (green line), 2M CaCl$_2$ (red line), and 2M MgCl$_2$ (blue line). (b-d) Differential absorption spectra of DMP solutions ($c_{DMP}$ = 0.2 M) containing different concentrations of Na$^+$, Ca$^{2+}$, and Mg$^{2+}$ ions. The dash-dotted lines in each panel represent the rescaled inverted linear absorption spectrum (thick black line in panel (a)) and the dashed lines the absorption spectrum of DMP interacting with ions. The dashed lines are the difference between the 2 M differential spectrum (blue lines in panels (b-d)) and the respective dash-dotted lines. Concentrations of 0.14$c_0$, 0.12 $c_{DMP}$, and 0.23$c_0$ are derived for DMP/Na$^+$, DMP/Ca$^{2+}$, and DMP/Mg$^{2+}$ contact ion pairs.

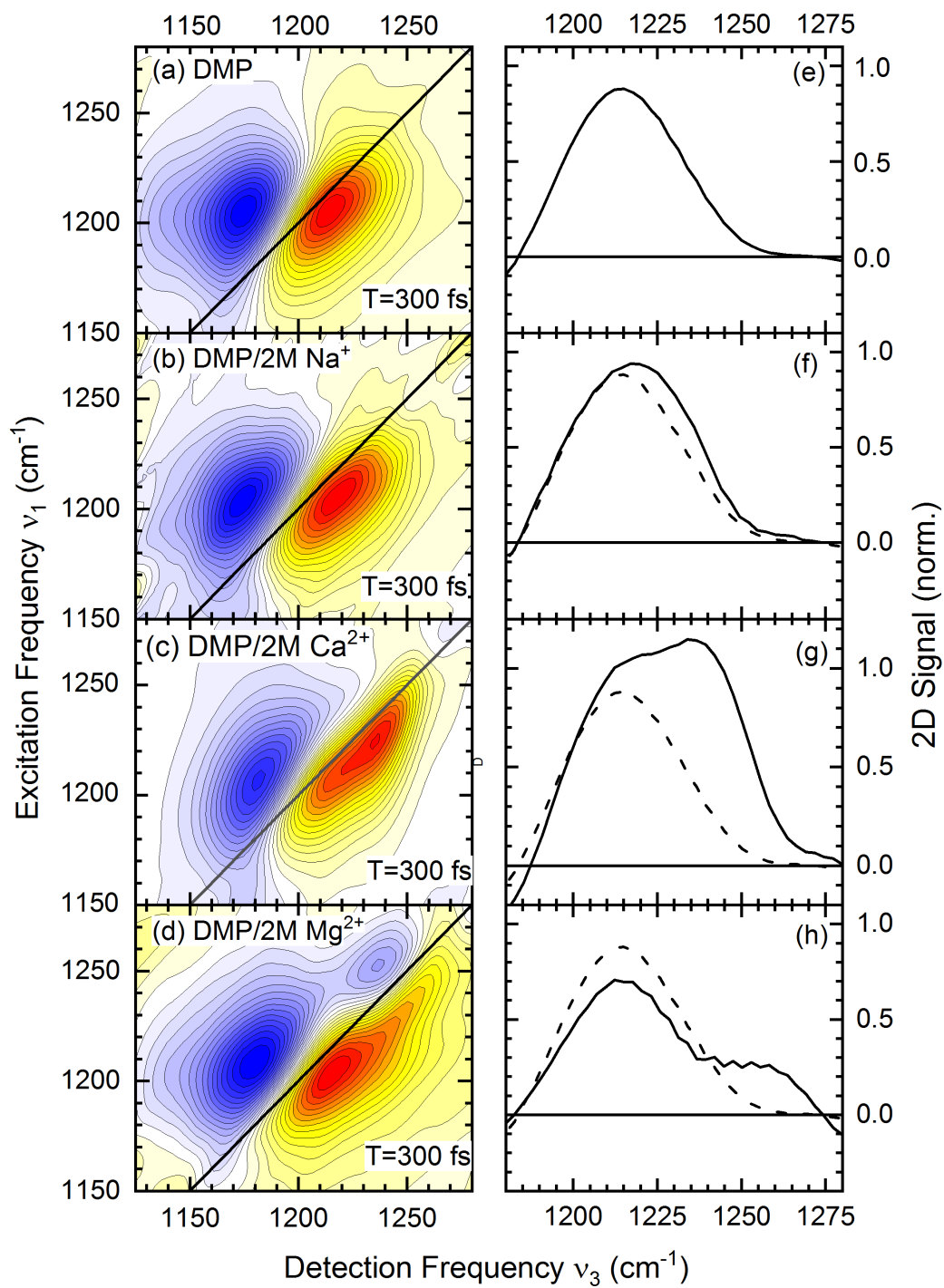

**Figure 2**. 2D-IR spectra of the asymmetric $(PO_2)^-$ stretching mode of (a) DMP in neat $H_2O$, (b) DMP in $H_2O$ with an excess of $Na^+$ ions, (c) DMP in $H_2O$ with an excess of $Ca^{2+}$ ions, and (d) DMP in $H_2O$ with an excess of $Mg^{2+}$ ions. The DMP and respective excess ion concentrations were 0.2 and 2 M. The absorptive 2D signal measured at a population time T=300 fs is plotted as a function of excitation frequency $\nu_1$ and detection frequency $\nu_3$. The yellow-red contours represent the signal due to the v=0→1 transition, the blue contours the signal on the v=1→2 transition. (e-h) Frequency cuts of the 2D-IR spectra along a diagonal crossing the maximum of the v=0→1 signals (solid lines). The dashed lines in panels (f-h) represent the diagonal cut of panel (e) for reference.

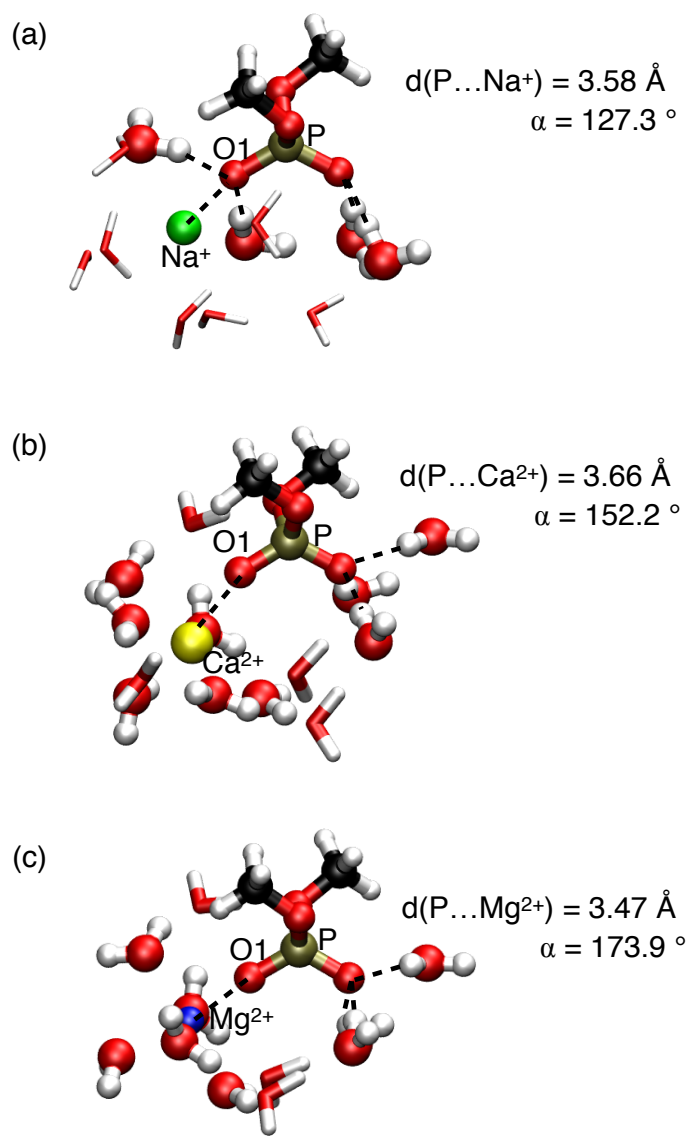

**Figure 3**. DMP(H$_2$O)$_N$M$^{x+}$ cluster minimum geometries (M$^{x+}$ = Na$^+$, Mg$^{2+}$: N = 11 and M$^{x+}$ = Ca$^{2+}$: N = 13). Water molecules in the first solvation shell around the ions and hydrogen bonded water molecules are in ball-and-stick representation, bridging water molecules in stick representation. (a) Prototypical solvation geometry of the DMP(H$_2$O)$_{11}$Na$^+$ cluster with α~127° and the Na$^+$ ion shown in green. The Na$^+$ ion intercalates into the tetrahedral hydrogen bond geometry around the O1 atom of the (PO$_2$)$^-$ group. (b) Prototypical solvation geometry realized in the DMP(H$_2$O)$_{13}$Ca$^{2+}$ cluster with α~152° and the Ca$^{2+}$ ion shown in yellow. The O1 atom of the (PO$_2$)$^-$ group takes the position of one of the water oxygen atoms in the first solvation geometry around the Ca$^{2+}$ ion. (c) Prototypical solvation geometry of the DMP(H$_2$O)$_{11}$Mg$^{2+}$ cluster for angle α~174° with the Mg$^{2+}$ ion shown in blue. The O1 atom of the (PO$_2$)$^-$ group takes the position of one of the water oxygen atoms in the octahedral solvation geometry around the Mg$^{2+}$ ion. Vibrational frequencies ν$_{AS}$(PO$_2$)$^-$ of (a) DMP(H$_2$O)$_{11}$Na$^+$, (b) DMP(H$_2$O)$_{13}$Ca$^{2+}$ and (c) DMP(H$_2$O)$_{11}$Mg$^{2+}$ are 1208, 1226 and 1248 cm$^{-1}$, respectively.

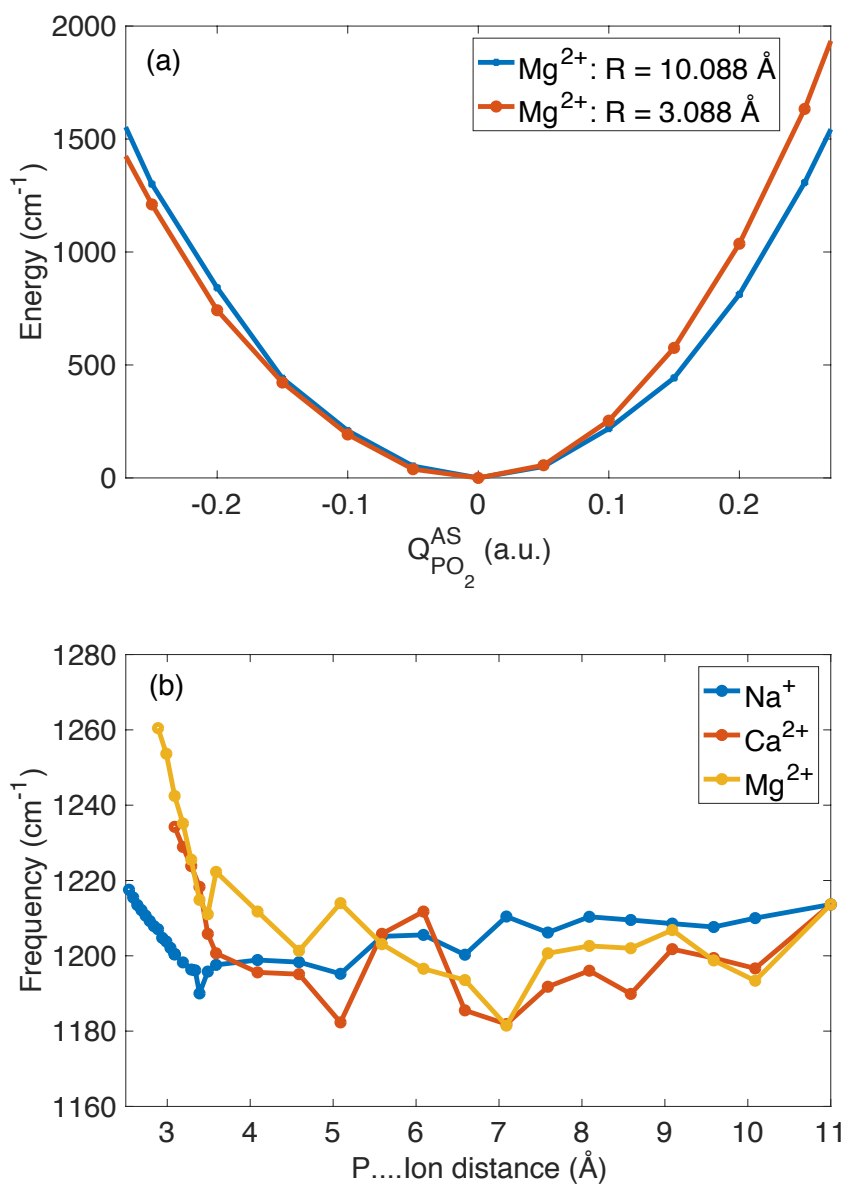

**Figure 4**. (a) Potential energy surfaces along the normal mode displacement vector of the asymmetric $(PO_2)^-$ stretching vibration ($Q^{AS}_{PO2}$) in the DMP$(H_2O)_{19}$Mg$^{2+}$ cluster for selected Mg$^{2+}$… P distances R. (b) Harmonic vibrational frequencies of $\nu_{AS}(PO_2)^-$ as a function of P…ion distance in the DMP$(H_2O)_{19}$M$^{x+}$ clusters (M$^{x+}$ = Na$^+$, Ca$^{2+}$ and Mg$^{2+}$).